%% file: ms.tex
\documentclass[12pt,preprint]{aastex}
\usepackage{amsmath}

\shorttitle{IRAS 17347$-$3139}
\shortauthors{Tafoya et al.}

\begin{document}


\title{A collimated, ionized bipolar structure and a high density torus in the young
	planetary nebula IRAS 17347$-$3139.}


\author{D. Tafoya\altaffilmark{1}, Y. G\'omez}
\affil{Centro de Radioastronom\'\i a y Astrof\'\i sica, Universidad Nacional Aut\'onoma de 
M\'exico, Apdo. Postal 3-72 (Xangari), CP 58089, Morelia, Michoac\'an, M\'exico}
\email{d.tafoya@astrosmo.unam.mx, y.gomez@astrosmo.unam.mx}

\author{N. A. Patel}
 \affil{Harvard-Smithsonian Center for Astrophysics, 60 Garden Street, Cambridge, MA 02138}
\email{npatel@cfa.harvard.edu}

\author{J. M. Torrelles}
\affil{Instituto de Ciencias del Espacio (CSIC)-IEEC, Facultad de F\'\i sica, Universitat 
de Barcelona, E-08028, Spain)}
\email{torrelles@ieec.fcr.es.}

\author{J. F. G\'omez, G. Anglada, L. F. Miranda}
\affil{Instituto Astrof\'\i sica Andaluc\'\i a, CSIC, E-18008 Granada, Spain}
\email{jfg@iaa.es, guillem@iaa.es, lfm@iaa.es}

\author{I. de Gregorio-Monsalvo}
\affil{European Southern Observatory, Alonso de C\'ordoba 3107, Vitacura, 
Casilla 19001, Santiago 19, Chile}
\email{idegrego@eso.org}


\altaffiltext{1}{Predoc Student at the Center for Astrophysics,
    60 Garden Street, Cambridge, MA 02138. Email:dtafoya@cfa.harvard.edu}

  
\begin{abstract}
\noindent
We present observations of continuum ($\lambda$ = 0.7, 1.3, 3.6 and 18 cm) and OH maser 
($\lambda$ = 18 cm) emission toward the young planetary nebula IRAS~17347$-$3139, which is one 
of the three planetary 
nebulae that are known to harbor water maser emission. From the continuum observations we show 
that the ionized shell of IRAS~17347$-$3139 consists of two main structures: one extended 
(size $\sim$1$\rlap{.}^{\prime\prime}$5) with bipolar morphology along PA=$-$30$^{\circ}$, 
elongated in the same direction as the lobes observed in the near-infrared images, and a 
central compact structure (size $\sim$0$\rlap{.}^{\prime\prime}$25) elongated in the direction perpendicular 
to the bipolar axis, coinciding with the equatorial dark lane observed in the near-infrared images. 
Our image at 1.3 cm suggests the presence of dense walls in the ionized bipolar lobes. We estimate 
for the central compact structure a value of the electron density at least $\sim$5 times higher than in the lobes. 
A high resolution image of this structure at 0.7 cm shows two peaks separated by about 
0$\rlap{.}^{\prime\prime}$13 (corresponding to 100-780 AU, using a distance range of 0.8$-$6 kpc). 
This emission is interpreted as originating in an ionized equatorial torus-like structure, from whose 
edges the water maser emission might be arising. We have detected weak OH~1612~MHz maser emission 
at $V_{\rm LSR}$~$\sim$~$-$70~km~s$^{-1}$ associated with IRAS 17347$-$3139. We derive a 3$\sigma$ 
upper limit of $<$ 35\% for the percentage of circularly polarized emission.  Within our primary beam, 
we detected additional OH~1612~MHz maser emission in the LSR velocity ranges $-$5 to $-24$ and $-$90 
to $-$123~km~s$^{-1}$, associated with the sources 2MASS J17380406$-$3138387 and OH 356.65$-$0.15, 
respectively.
\end{abstract}


\keywords{planetary nebulae: general --- stars: AGB and post-AGB --- stars: 
winds, outflows --- stars: mass loss --- stars: evolution --- radio continuum,
masers: bipolar --- planetary nebulae: individual(IRAS~17347$-$3139)}

\section{Introduction}

The study of transition objects from the asymptotic giant branch (AGB) to the planetary nebula (PN) 
phase is very important to understand the processes by which low and intermediate mass stars evolve. 
It has been observed that planetary nebulae (PNe) display a large variety of morphologies, including 
bipolar or multipolar structures (Balick 1987; Schwarz, Corradi, \& Melnick 1992; Manchado et al. 
1996). However, it is not well understood how they develop such morphologies. Given that the transition 
phase occurs in a very short time scale of $\sim$1000 years (Kwok 1993), only a few objects are 
expected to be in this evolutionary stage, making the observational study of the physical conditions 
under which they evolve a difficult task. A simple model which in general explains the development of
 bipolar morphologies in PNe is the generalized interacting stellar winds (GISW) model (Kahn \& West 
1985; Balick 1987; Icke 1988, Mellema et al. 1991). This model assumes that the ``superwind'', expelled 
during the AGB phase, produces a circumstellar envelope (CSE) which has a latitude-dependent density profile,
 with an enhancement in the equatorial region and decreasing monotonically toward the poles. Subsequently,
 the slow massive wind is replaced by a fast tenuous wind; the latter interacts hydrodynamically with
 the former, resulting in the creation of the bipolar lobes (Mellema et al. 1991, Frank et al. 1993, 
Garc\'\i a-Segura et al. 1999, Balick \& Frank 2002). 

High angular resolution and sensitive images of PNe obtained with the Hubble Space Telescope (HST) have 
revealed collimated structures whose formation cannot be explained by the GISW model (Miranda \& Solf 1992; 
Sahai \& Trauger 1998). The presence of a companion, collimated outflows (e.g.  Sahai 
\& Trauger 1998; Soker \& Rappaport 2000; Vel\'azquez et al. 2007) or magnetic fields (e.g. 
Garc\'\i a-Segura et al. 1999), are required in most cases to explain the formation of such collimated 
structures. Nonetheless, the existence and study of a disk or an equatorial density enhancement in the CSE 
is considered a key ingredient to understand the processes that form bipolar 
lobes. A detailed study of particular PNe can provide crucial information about the physical conditions 
under which they develop their morphologies, and help to determine the relevance of the different 
shaping mechanisms proposed.

IRAS~17347$-$3139 is a young PN with a clear bipolar morphology, as revealed by the near-infrared images 
(de~Gregorio-Monsalvo~et~al. 2004 [hereafter dGM04], S\'anchez-Contreras et al. 2006; Sahai et al. 2007). 
The lobes show an extent of $\sim$4$^{\prime\prime}$, separated by a dark lane, which probably is tracing 
a dense dusty equatorial region. S\'anchez-Contreras et al. (2006) suggest that the limb brightened 
appearance of the lobes could be indicating the presence of bubble-like structures with dense walls 
and tenuous interiors, presumably excavated by jet-like winds.
   
dGM04 detected water masers arising from this young planetary nebula. Up to now, only other two PNe are
 known to exhibit water maser emission (Miranda et al. 2001; G\'omez et al. 2008). Since the water maser
 emission is expected to last for a very short period after the intense mass-loss rate stops, at the 
end of the AGB phase ($\sim$ 100 yr, G\'omez, Moran \& Rodr\'\i guez 1990), the detection of this emission
 suggests that these stars have entered the PN phase only some decades ago, making this objects good 
candidates to study the early stages of PN formation. On the other hand, the nature 
of the radio continuum emission in IRAS~17347$-$3139 has been discussed by dGM04 and G\'omez et al. 2005 
(hereafter G05). These authors showed that the flux density of IRAS~17347$-$3139 rises with frequency, 
deriving a spectral index $\alpha $~$\simeq$~0.7 ($S_{\nu}\propto \nu^{\alpha}$), between 4.9 and 22 GHz, 
which was interpreted in terms of free-free emission from an ionized nebula. Moreover, G05 found that the 
radio continuum flux density is increasing rapidly with time. They estimated a dynamic time scale for
 the ionized envelope of $\sim$ 100 yr, supporting the idea that this star entered the PN phase only
 some decades ago.   

OH maser emission at 1612 MHz toward IRAS~17347$-$3139 was first reported by Zijlstra et al. (1989). 
Recently, Szymczak and G\'erard (2004) presented single-dish polarimetric observations of OH masers 
toward this source. However, the association of this emission with the PN is uncertain due to the 
low angular resolution of their observations.

In order to clarify some questions originated in previous works, and to further investigate the PN 
IRAS~17347$-$3139, we have carried out high sensitivity and angular resolution continuum and OH
 maser observations with the Very Large Array (VLA). This work is structured as follows: In \S~2 
we describe the new observations that allowed us to image with higher sensitivity and higher angular 
resolution the ionized envelope of this source. The results are presented in \S~3, the analysis of the 
data is discussed in  \S~4, and the conclusions are given in \S~5.

\section{Observations and data reduction}

\noindent
On 2005 January 27, we used the VLA of the NRAO\footnote[2]{The National Radio 
Astronomy Observatory is a facility of the National Science Foundation operated under 
cooperative agreement by Associated Universities, Inc.} in the hybrid configuration BnA, 
to carry out continuum observations of IRAS~17347$-$3139 at frequencies 8.46, 22.46 and 
43.34 GHz ($\lambda$ = 3.6, 1.3 and 0.7 cm, respectively). We used 2 IF's covering 
a total bandwidth of 100~MHz with two circular polarizations. At 0.7 and 1.3 cm we used 
the fast switching mode, changing from source to phase calibrator every 80 seconds in 
order to correct for the quick variations in the troposphere. As the phase tracking center, 
we used the position of the peak of the radio continuum  reported by dGM04: 
$\alpha$(J2000.0)~=~17$^{\rm h}$38$^{\rm m}$00$\rlap{.}^{\rm s}$586, 
$\delta$(J2000.0)~=~$-$31$^{\circ}$40$^{\prime}$55$\rlap{.}^{\prime\prime}$67. 
The source J1335+305 (3C~286) was used as the flux calibrator while J1744$-$312 was the 
phase calibrator. Table \ref{table_1} lists the flux densities of the calibrators at the 
different frequencies. The total time on source was 0.6, 0.5 and 1.1 hrs at $\lambda$ = 3.6, 
1.3 and 0.7 cm, respectively.

We also carried out, with the VLA-BnA configuration, spectral line observations of four OH 
transitions with rest frequencies: 1612, 1665, 1667 and 1720 MHz. The 1665 and 1667 MHz 
observations were carried out on 2005 January 28, while the 1612 and 1720 MHz observations were carried
 out on 2005 January 29. For each transition we observed both right and left circular polarizations 
(RCP and LCP, respectively). We sampled 256 channels in a total bandwidth of 1.5625 MHz, centered
 at $V_{\rm LSR}$=$-$40 km~s$^{-1}$, 
resulting in a spectral resolution of 6.1035 kHz ($\simeq$ 1.1 km~s$^{-1}$). In addition, for 
the 1612 MHz transition, we used a narrow-band filter to avoid radio frequency interference due 
to the Iridium satellites. The sources 3C~286 and J1751$-$253 were the flux and phase calibrators, 
respectively (see flux densities in Table \ref{table_1}). From the spectral data of each 
transition we have obtained a continuum data set by averaging channels free of maser emission. 
The four continuum data sets were calibrated separately. Subsequently, during the imaging process, 
they were concatenated to produce a single continuum data set.

The calibration and data reduction were carried out using the Astronomical Image Processing
 System (AIPS) of the NRAO. We followed the standard procedures for reducing high-frequency data 
outlined in Appendix D of the AIPS Cookbook (2007). Since 3C~286 is resolved at some of the observation 
frequencies, we used an image model for this source in the calibration process. The data for the 
continuum observations were self-calibrated in phase (except at $\lambda$=18 cm), then Fourier transformed, 
weighted, and CLEANed to generate the final images.

\section{Results}

\subsection{Radio continuum}

\begin{figure}
\begin{center}
\vspace{3cm}
\includegraphics[angle=-90,scale=0.7]{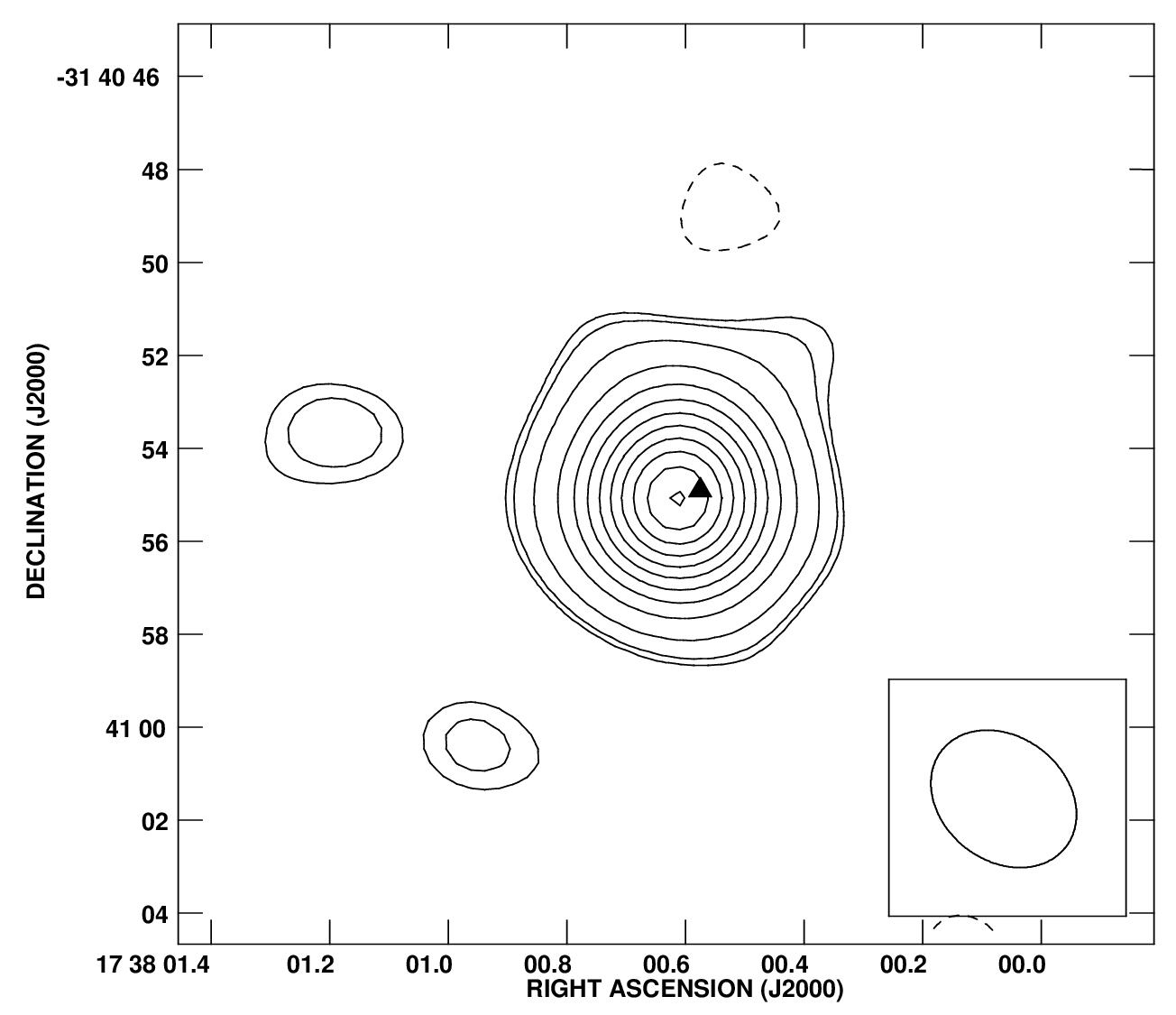}
 \caption{VLA image of IRAS 17347$-$3139 at wavelength 18 cm using a ROBUST weight parameter 
$= 0$. Contours are $-$3.7, 3.7, 5, 10, 20, 30, 40, 50, 60, 70, 80, 90, and 99 per cent
 of 2.4$\times 10^{-2}$ Jy/beam, the peak value of the brightness. The first contours are -3 and 3 
times the  $rms$ noise of the image, 2.9 $\times 10^{-4}$ Jy/beam.  The synthesized beam is shown
 in the bottom right corner, and its size is $3\rlap{.}^{\prime\prime}38 \times 2\rlap{.}^{\prime\prime}67$ 
(PA = $52^\circ$). The position of the OH~1612 MHz maser associated to IRAS~17347$-$3139 is show 
as a filled triangle.  
\label{fig_1}}
 \end{center}
\end{figure}

\begin{figure*}
\begin{center}
\vspace{3cm}
\includegraphics[angle=-90,scale=0.7]{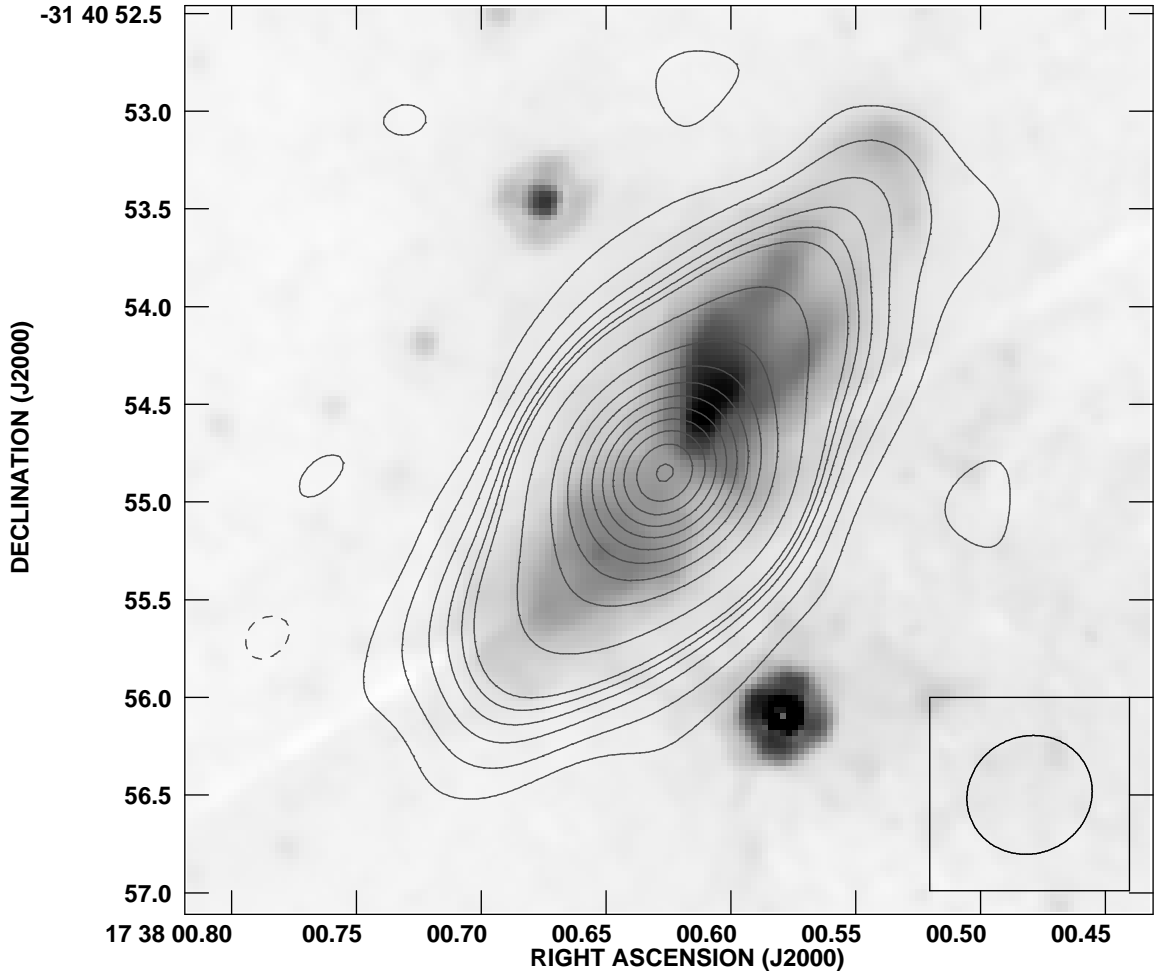}
 \caption{Contours: VLA image of IRAS 17347$-$3139 at wavelength 3.6 cm using a ROBUST weight 
parameter $= 0$. Contours are $-$0.31, 0.31, 1, 2, 3, 4, 5, 10, 20, 30, 40, 50, 60, 70, 80, 90 
and 99 per cent of 7~$\times 10^{-2}$~Jy/beam. The value of the first contour 
is 3 times the $rms$ noise of the image, 7.2~$\times 10^{-5}$~Jy/beam. The synthesized beam is 
shown in the bottom right corner, and its size is $0\rlap{.}^{\prime\prime}65 \times 
0\rlap{.}^{\prime\prime}60$ (PA = $-65^\circ$). 
Grey scale: HST IR image of IRAS 17347$-$3139 at 1.1 $\mu$m obtained with the NIC1 in the F110W 
filter. In order to make coincide the nominal position of the IR emission with the peak of 
the radio continuum, the IR image has been shifted $\sim 0\rlap{.}^{\prime\prime}5$ to the southeast.
\label{fig_2}}
 \end{center}
\end{figure*}

\begin{figure*}
\begin{center}
\vspace{3cm}
\scalebox{0.7}{\includegraphics[angle=-90]{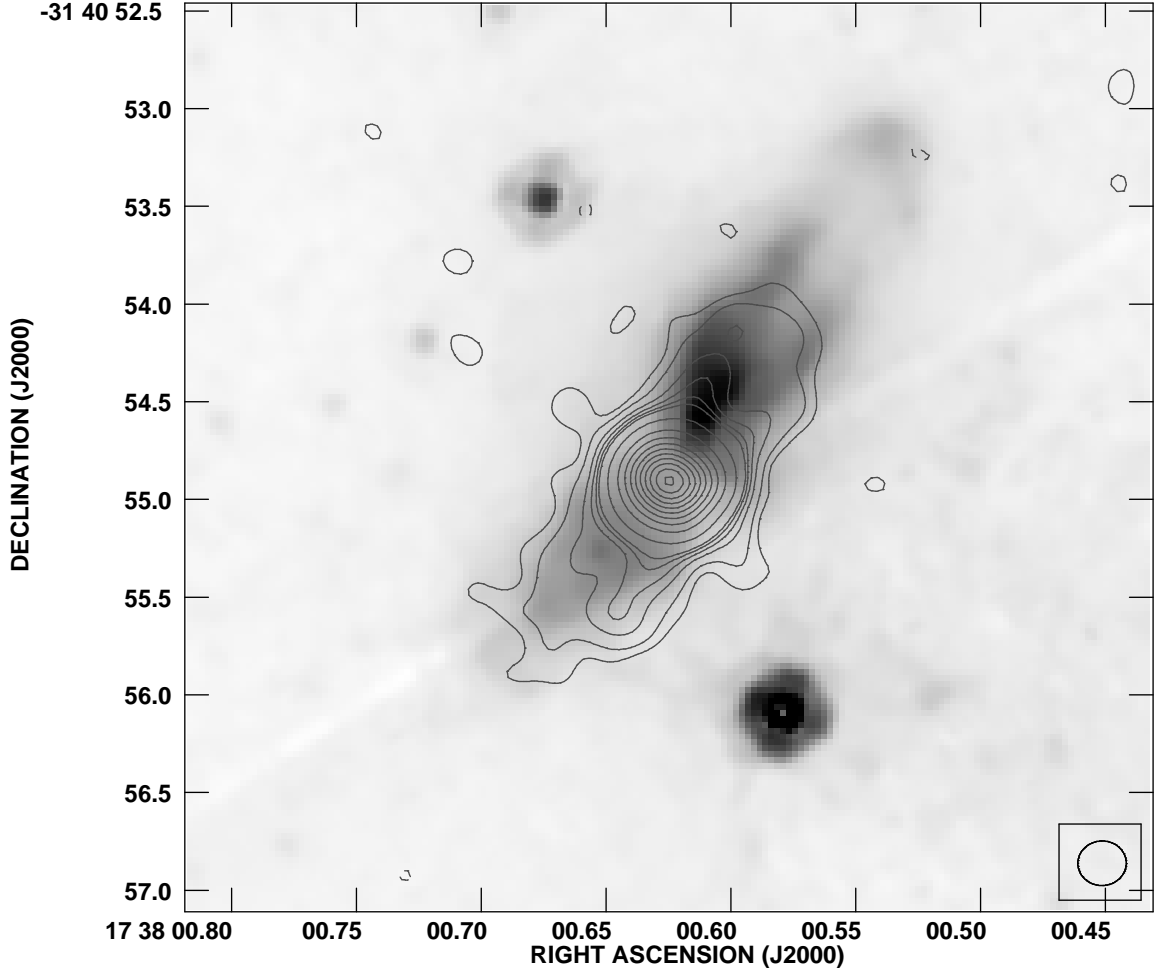}}
 \caption{Contours: VLA image of IRAS 17347$-$3139 at wavelength 1.3 cm using a ROBUST weight parameter 
$= 0$. Contours are $-$0.53, 0.53, 1, 2, 3, 4, 5, 10, 20, 30, 40, 50, 60, 70, 80, 90 and 
99 per cent of 1.1 $\times 10^{-1}$ Jy/beam, the peak value of the brightness. The value of the 
first contour is 3 times the $rms$ noise of the image, 2.1$\times 10^{-4}$ Jy/beam. The synthesized 
beam is shown in the bottom right corner, and its size is $0\rlap{.}^{\prime\prime}25 \times 0\rlap{.}^{\prime\prime}23$ (PA = $-88^\circ$). 
Grey scale: HST IR image of IRAS 17347$-$3139 at 1.1 $\mu$m obtained with the NIC1 in the F110W 
filter.
\label{fig_3}}
 \end{center}
\end{figure*}

Since the first detection by Zijlstra et al. (1989), it is known that IRAS~17347$-$3139 
presents radio continuum emission. However, there were no radio images of this 
source reported in the literature. From our observations, we have obtained high sensitivity 
and high angular resolution interferometric images of IRAS~17347$-$3139 at wavelengths 18, 
3.6, 1.3 and 0.7 cm (Figs. \ref{fig_1}, \ref{fig_2}, \ref{fig_3} and \ref{fig_4}). This is the first
 time that the radio emission from this PN has been resolved spatially.

At 3.6 and 1.3 cm (Figures \ref{fig_2} and \ref{fig_3}, respectively) the radio continuum emission shows a
 bright central region and a fainter extended structure elongated in the northwest-southeast direction 
(P.A.~=~$-30^\circ$). Particularly, in the image at 1.3 cm, the extended emission shows a double horn 
structure in the nortwest-southeast direction with an opening angle of about 30$^{\circ}$. 
This structure also appears in the near-infrared emission (see also dGM04; S\'anchez-Contreras et al. 
2006, Sahai et al. 2007). The bright emission of the central region at 1.3 cm seems slightly 
elongated in the direction perpendicular to the extended structure. This elongation of the central 
region is more clearly seen in the image at 0.7 cm (Fig. 4). 

We fitted two bi-dimensional Gaussian components to the emission observed at 3.6 cm, and also to the 
emission at 1.3 cm, in order to estimate the size and orientation of both the compact (central region)
 and extended emission. At 18 and 0.7 cm only one Gaussian component was fitted\footnote[3]{We tried to 
fit two Gaussian components to the emission at 0.7 cm but, due to the low signal-to-noise ratio of the
 extended structure, the fitting 
was doubtful.}. For the fitting process, we used images reconstructed with the ROBUST parameter 
(Briggs 1995) set to +5 in order to recover as much faint extended emission as possible, then we used
 the task JMFIT 
of AIPS to fit the Gaussians. The results of the fitting are shown in Table \ref{table_2}. From 
these images (ROBUST $=+5$), we also measured the total flux density of the source within a box 
containing the whole emission (Table \ref{table_3}). We notice that the total flux density at 1.3, 3.6 
and 18 cm is compatible with the sum of the flux densities of the fitted Gaussian components. 
At 0.7 cm, the flux of the fitted Gaussian is lower than the total flux of the source. This is 
probably due to the presence of extended faint emission not fitted by a single Gaussian component. 
Therefore, in Table \ref{table_2}, we attribute the residual flux density from the Gaussian 
fitting at 0.7 cm to the extended structure.  

In Figure \ref{fig_5} we compare our measurements of the total flux density with those of previous
 works and epochs. We noticed that the flux densities that we obtained follow an increasing trend
 with time, just as it was found previously by G05. A possible explanation of the increase in flux density 
could be the expansion of the ionized nebula (given that the spectral index does not vary significantly 
between the different epochs). From our observations we estimate a spectral index in the frequency range 
from 1.6 to 43 GHz of $\simeq$~0.81, which is similar to the values found by dGM04 and G05. Using the 
values of the flux density at 3.6 cm from our observations and from those of dGM04 and G05, assuming 
a constant expansion velocity, we estimate a kinematical age for the ionized nebula of about 100 years.
 This value is consistent with the result obtained by G05.

In Table \ref{table_3} we list the continuum peak positions of the source at the different 
frequencies. At $\nu =$ 8.46, 22.46 and 43.34 GHz (for which the source is better resolved), they 
coincide within $0\rlap{.}^{\prime\prime}06$ from each other. When this position is compared with 
the nominal infrared position of the nebula obtained from the HST image, the difference is of about 
0$\rlap{.}^{\prime\prime}$5. We note, however, that the position we have
measured for the peak emission differs by about 0$\rlap{.}^{\prime\prime}$9 from that 
given by dGM04, that was obtained from data with a lower angular resolution. Since the newer specific 
procedures to calibrate high frequency data were not used by dGM04, we attribute this difference to 
their absolute positional error, that appears to be $\sim 1''$.

\subsection{OH Maser Emission}

Among the four OH maser transitions observed at $\lambda \sim 18$ cm, we only detected the line at 1612 MHz 
toward IRAS 17347$-$3139. In Table \ref{table_4} we give the parameters of the OH maser line detected. The 
OH~1612 MHz maser emission is slightly displaced to the north-west from the peak of the radio continum at 18 
cm ($\sim$ 0$\rlap{.}^{\prime\prime}$5, Figure \ref{fig_1}). In Figure 
\ref{fig_6} (left panel) we present the spectra of the RCP and LCP ($S_{\rm RCP}$ and $S_{\rm LCP}$, 
respectively), as well as the total flux density (Stokes I=[$S_{\rm RCP}$+$S_{\rm LCP}$]/2) and the 
circular polarization (Stokes V=[$S_{\rm RCP}$$-$$S_{\rm LCP}$]/2) spectra. In both polarizations there
 is only one feature at velocity $V_{\rm LSR} \simeq -70$ km~s$^{-1}$ with a flux density of $\simeq$~40 
mJy. From the Stokes V spectrum we do not find evidence of the presence of circular polarized emission.
 We estimate a 3-$\sigma$ upper limit of $<$~35\% for the percentage of circularly polarized emission (m$_{c}$=$|$V$|$/I).

In addition to the OH 1612 MHz maser emission coming from IRAS 17347$-$3139, we also detected this line 
from two other positions that are located within the primary beam of our observations, which is 
30$^{\prime}$ at this frequency. One $\sim$ 2$\rlap{.}^{\prime}$5 northeast from IRAS~17347$-$3139 and 
the other is $\sim$ 11$^{\prime}$ south from this PN (see Table \ref{table_4}). 
The maser emission located toward the northeast appears 
in the velocity range from $-$125 to $-$95 km~s$^{-1}$ (right panel of Figure \ref{fig_6}), which is 
very similar to 
the velocity range of the OH maser emission that Zijlstra et al. (1989) reported to be associated with 
IRAS~17347$-$3139. However, from our observations we found that the position of the OH maser emission in this 
velocity range coincides with the 2MASS source J17380406$-$3138387. On the other hand, the emission located 
toward the south of the position of IRAS~17347$-$3139 appears in the velocity range from $-$23 to 
5 km~s$^{-1}$ (right panel of Figure \ref{fig_6}). This emission is associated with the source 
OH~356.65$-$0.15 
and was already reported by Bowers \& Knapp (1989). Recently Szymczak \& G\'erard (2004) reported OH maser
 emission toward IRAS 17347$-$3139 in the same velocity range. However, due to the low angular resolution
 of their observations, probably they were contaminated by emission from the source OH~356.65$-$0.15.

\medskip
\section{Discussion}


The spectrum of the free-free emission of an ionized region depends on the 
geometry, as well as on the electronic density and temperature 
distributions inside the region. For the simplest case of an isothermal, 
homogeneous ionized region the value of the spectral index, $\alpha$ 
(where $S_\nu \propto \nu^\alpha$), ranges from +2 (at low frequencies, 
where all the emission is optically thick) to $-0.1$ (at high frequencies, 
where all the emission is optically thin). The maximum flux density is 
reached at the ``turnover frequency'', $\nu_{m}$, where the optical depth 
is of the order of unity and the spectrum becomes flat. In general, for a 
non-homogeneous ionized region, the spectral index of the free-free 
emission is the result of contributions from lines of sight with different 
optical depths. For the case of an isothermal, ionized region where the 
electron density goes as the inverse square of the radius ($n_{e} \propto 
r^{-2}$), the emission from the inner part of the region would be 
optically thick, while that from the outer parts would be optically thin, 
resulting in a constant value of the spectral index $\alpha=+0.6$ over a 
wide range of frequencies (Panagia \& Felli 1974;  Olnon 1974; Reynolds 
1986). If the ionized region is truncated at an inner radius $r_0$, there 
is a turnover frequency, where all the emission is optically thin and the 
spectrum becomes flat. These properties are true both for a spherically 
symmetric region and for a biconical (constant opening angle) region. This 
is also true for a constant velocity ionized spherical (or biconical) wind 
since in this case the electron density also decreases as the square of 
the radius (if the ionized fraction remains constant). If the density 
decreases steeply (e.g., in an accelerating wind), then the value of 
spectral index would be higher than +0.6, while a smaller value of the 
spectral index would indicate a flatter density distribution (e.g., in a 
decelerating wind).

The spectral index of the radio emission from IRAS~17347$-$3139 is
$\alpha\simeq0.81$ (Fig. \ref{fig_5}) in the range of frequencies from
1.6 to 43 GHz. This can be interpreted in terms of an ionized region in
which the electron density decreases as $n_{e} \propto r^{-2.3}$. This
density distribution could correspond to a wind with an almost constant velocity 
($v \propto r^{0.3}$). dGM04 and G05 discussed the possibility of the
presence of an ionized wind in IRAS~17347$-$3139, but the mass-loss rate
derived from that assumption, $\dot M_i \simeq
10^{-4}$~($D$/kpc)$^{3/2}~M_{\odot}$~yr$^{-1}$, is far larger than the
values observed in central stars of PNe and pre-PNe ($\leq$ 10$^{-7}$
$M_{\odot}$ yr$^{-1}$; Patriarchi \& Perinotto 1991;  Vassiliadis \& Wood
1994). As a result, these authors interpreted the emission as being
arising from a recently ionized nebula.

Our new observations provide further insights on the nature of the radio
emission of IRAS~17347$-$3139.  As metioned in \S~3.1, our high angular
resolution images reveal the presence of two structures elongated in more
or less perpendicular directions. In fact, the Gaussian fitting to the
continuum emission of IRAS~17347$-$3139 (\S~3.1) shows that the central
compact structure is elongated in the direction with P.A.$\simeq$
50$^{\circ}$, while the extended structure is elongated in the direction
with P.A.$\simeq$ $-$30$^{\circ}$. These directions correspond to those of
the dark lane and the bipolar bright lobes observed in the IR images,
respectively (see Figs. 2, 3, and 4). This alignment suggests that the radio 
continuum emission could be arising from two different components: an equatorial ionized
torus-like structure, and two ionized bipolar lobes. To test this
hypothesis, we analyzed the two components separately.

\medskip
\subsection{The Extended Ionized Emission}

Since, as mentioned above, the spectrum of the free-free emission from an
ionized region depends on the geometry and physical properties ($T_e$ and
$n_e$ distributions), we can use this information to infer some of the
properties of the extended structure. In Figure \ref{fig_7} (left panel)
we have plotted the spectrum of the extended component, resulting from the
values of the flux density obtained from the Gaussian fitting (\S~3.1).  
From 8.4~GHz to 43~GHz, the spectrum is flat, indicating that the turnover
frequency is $\nu_m < 8.4$~GHz. There is only one measurement at
frequencies lower than 8.4 GHz and, therefore, only a lower limit, $\alpha
> +0.5$, can be obtained for the spectral index in the partially opaque
regime. High angular resolution observations at 6 cm would be useful to
constrain the value of this spectral index. Our observations show that the
geometry of this component is not spherical, but has an aperture angle
of $\theta_0\simeq 30^{\circ}$; therefore, it can be better
described as a collimated ionized region. If we assume that the aperture
angle is constant as a function of the distance to the central star
(biconical structure), and that the inclination of this structure with
respect to the plane of the sky is small, adopting an spectral index
$\alpha \simeq +0.5$, a turnover frequency $\nu_m \simeq 8$~GHz, and an
electronic temperature $T_{e}=10^{4}$~K, we derive an inner radius 
$r_0 \simeq 0\rlap.''3$, and a density profile $n_{e} \propto r^{-1.9}$
(eqs. 15 and 18 from Reynolds 1986). A value of the spectral index higher
than +0.5 would result in a density distribution decreasing steeply.

The electron density at radius $r_{0}$ is independent of $\alpha$, and can
be estimated from equation~(13)  of Reynolds (1986) as:
\begin{equation}\label{equation_1}
\left[\frac{n_0}{\rm cm^{-3}}\right] = 1.12\times 10^{3}\left[\frac{w_0}{\rm arcsec}\right]^{-0.5}
\left[\frac{T_e}{\rm K}\right]^{0.675} 
\left[\frac{\nu_{m}}{\rm GHz}\right]^{1.05}
\left[\frac{D}{\rm pc}\right]^{-0.5},
\end{equation}
 where $w_0=\theta_0 r_0/2 \simeq 0\rlap.''08$) is the width of the ionized
cone at radius $r_{0}$, and $D$ is the distance to the source. For a
distance range from 6 to 0.8 kpc (G05), we derive an electron density at
the base of the lobes of $n_0=2$-$6\times10^{5}$~cm$^{-3}$. Although this
is the maximum value of the density in the lobes, and the average value
would be smaller (e.g., at $r=1''$ the electron density would be ten times
smaller than $n_0$), this high value of the electron density in the
bipolar lobes supports the idea that this is a very young PN, and that the
double horn structure seen at 1.3 cm (\S 3.1) could be tracing high
density walls.


If we further assume that the extended emission arises from a biconical ionized 
wind with an expansion velocity of 1000~km~s$^{-1}$, using equation~(19) from 
Reynolds (1986), we estimate a mass-loss rate of 
$\dot{M}_{i}$~$\simeq$~1.3$\times$10$^{-5}$($D$/kpc)$^{3/2}$. This value is, 
once again, much larger than those observed in the stellar winds of other PN, 
favoring the idea that the emission arises from an ionized nebula. Nonetheless, 
it is worth noting the relatively high degree of collimation of this emission. 
Moreover, in the image at 3.6 cm the presence of a subtle point-symmetry is 
suggested. In the northern lobe this emission extends all the ways toward the 
tip where there is a bow-shaped structure. This morphology could be indicating 
that the extended ionized regions were excavated by a collimated wind (see \S4.3). 

\medskip
\subsection{The Central Region: a High Density Ionized Torus?}
 
In Figure \ref{fig_7} (right panel) we show the spectral energy
distribution (SED) of the central compact region from 8 to 43~GHz. For
this range of frequencies the spectral index is $\alpha \simeq 1$, and
seems to become shallower at higher frequencies. The value of the spectral
index of this component indicates that the geometry or the density
distribution of the ionized gas are different than in the extended
component. Also, the fact that the flux density apparently continues to
increase for frequencies as high as 43 GHz indicates that part of this
component is still optically thick at such frequencies.  Considering that
the optical depth is of the order of unity near the turnover frequency,
$\nu_{m} \geq$ 43~GHz, assuming a constant electronic temperature,
$T_{e}$=10$^{4}$~K, and that the size of this component along the line of
sight is similar to the width obtained from the Gaussian fit ($\sim$
0$\rlap{.}^{\prime\prime}$25; Table \ref{table_2}), we obtain a lower
limit for the electron density of $n_{e} \geq 1$-$3\times10^6$~cm$^{-3}$
for the distance range from 6 to 0.8 kpc. This value for the density is 5
times larger than the maximum value found for the extended emission.  
Considering this higher value of the density, and the elongation of this
component in the direction perpendicular to the lobes of the extended
emission, we suggest that the radio continuum emission is tracing the
inner ionized regions of an equatorial torus-like structure, which is
observed in the IR image as a dark lane, or waist (see Figs. 2, 3, and 4).

To further confirm this suggestion, we made images at 0.7 cm using only
the baselines longer than 500 k$\lambda$ (and ROBUST parameter = 0) to
improve the angular resolution (Fig.~\ref{fig_8}).  In this image a
double-peaked structure is observed, with the peaks symmetrically
separated with respect to the position of the radio continuum peak
measured at other frequencies. The separation between the peaks ($\simeq$
0$\rlap{.}^{\prime\prime}$13) corresponds to 100$-$780 AU for a distance
range of 0.8$-$6 kpc. A similar double peaked structure was observed in the
PN NGC~2440 by V\'azquez et al. (1999).
 From their observations of radio continuum and recombination lines, these
authors inferred the presence of large extinction toward the central
region, as well as a rotating, ionized toroid, roughly perpendicular to
the bipolar lobes. In the case of IRAS~17347$-$3139 the two radio
continuum peaks appear toward the dark lane observed in the IR images,
and they are aligned in the direction perpendicular to the axis of the
lobes, suggesting that the emission is also arising from an ionized
torus-like region.

The water masers detected toward IRAS~17347$-$3139 by dGM04 appear distributed 
in an elliptical structure with its center shifted
$\sim$~0$\rlap{.}^{\prime\prime}$15 from the peak of the radio continuum. These 
authors suggested that the peak of the radio continuum
could be tracing one stellar component in a binary system while the masers would 
be associated with a companion star. The angular
resolution of their observations ($0\rlap.''81\times0\rlap.''26$; PA = $26^\circ$) 
did not allow to confirm this suggestion. From our
observations we were able to compare the distribution of water masers relative to
 the radio continuum emission with higher angular
resolution. In Figure \ref{fig_8}, we have superimposed the water maser emission 
found by dGM04 on the radio continuum emission at 0.7 cm from
our observations. In order to do this, we have shifted the positions of the 
observations of dGM04, so that the position of the continuum
peak emission of their observations coincides with that of our observations at 
$\lambda \simeq$ 1.3 cm.  From this Figure there is no
clear evidence of the presence of a secondary companion associated with the maser
 emission as suggested by dGM04, although this
possibility cannot be completely ruled out. If we consider that the emission at 
0.7 cm is tracing an ionized torus around the central
star(s), the relative positions of the maser and the continuum emission suggest 
that the water masers arise from the outer parts of the ionized torus.

\medskip
\subsection{Collimated winds and tori in PNe}
 
It is now well known that several proto-PNe and PNe show the presence of collimated 
structures which often have point-symmetric morphology (Schwarz, Corradi, \& Melnick 1992; 
Sahai \& Trauger 1998; Balick \& Frank 2002). It has been 
suggested that these structures are created by collimated winds or jets. Lim \& Kwok 
(2000, 2003) detected collimated radio emission in the core of the PN M2-9 which they 
interpreted as arising in an ionized jet. More recently, Lee et al. (2007) found optically 
thick radio cores in narrow-waist bipolar nebulae. They suggest that the radio continuum 
emission is arising in collimated ionized winds, which would be responsible for the 
shaping of the PNe. 

As mentioned in \S4.1, from our radio continuum observations, we found that the ionized 
extended component of the PN IRAS~17347$-$3139 shows a relatively high degree of collimation and 
that its spectral index is consistent with that of an ionized wind. We also note that the image
 of IRAS 17347$-$3139 at 3.6 cm (Fig. 2) suggests that the extended emission shows a subtle 
point-symmetry. Furthermore, a closely look of the emission at 1.3 and 0.7 cm (Figs. 3 and 4) 
reveals that it also shows point-symmetry which is consistent with the point-symmetry observed 
at 3.6 cm. In particular, the emission at 0.7 cm has two faint bumps, one toward the north and
 the other toward the south. These point-symmetric morphologies have been observed in several 
proto-PNe and PNe (Corradi et al. 1993; Guerrero et al. 1999; Miranda et al. 2001). 
They have been interpreted as the result of the presence of precessing jet-like winds excavating 
the slowly expanding envelope ejected during the AGB phase (Miranda et al 2001; Volk et al. 
2007). Consequently, based on its morphological structure, we suggest that part of the emission 
of the extended component of IRAS 17347$-$3139 could be arising in a precessing ionized wind. 
Radio recombination line observations could be useful to probe the kinematics of the ionized 
nebula and confirm the presence of an ionized wind in this source. 
 
The collimated winds have been succesful at explaining the formation of point-symmetric 
morphologies in proto-PNe and PNe (Garc\'\i a-Segura 1997; Vel\'azquez et al. 2007). However, 
the mechanims responsible for the launching and collimation of such winds remain poorly 
understood. It has been observed that the pre-PNe and PNe that present collimated winds often 
also show the presence of an equatorial torus (Miranda et al. 2001; Sahai et al.2005; Uscanga 
et al. 2008). It has been proposed that the equatorial tori somehow could be related to the 
formation of the collimated winds (Mellema et al. 1991, Frank et al. 1993; Huggins 2007). 
In the GISW model, the torus channels the fast wind toward the polar regions, however it 
cannot collimate it (Garc\'\i a-Segura et al. 1999). Our observations reveal that a torus
 could be present in the equatorial region of IRAS~17347$-$3139 (see \S4.2), however, given 
its narrow waist and the possible presence of a precessing collimated ionized wind, the GISW 
model cannot explain the shape of this PN.

One of the mechanisms proposed to produce collimated outflows assumes that the fast wind is 
magnetized. Additionally, in the presence of a binary companion, the wind could undergo precession, 
producing a point-symmetric morphology (Garc\'\i a-Segura 1997). However, in these models, while the 
presence of an equatorial torus is possible, it is not indispensable for the collimation of the 
jets. On the other hand, Nordhaus \& Blackman (2006) proposed that a secondary companion could 
spiral-in the AGB circumstellar envelope enhancing the magnetic field by dynamo action. During the 
spiral-in process, an equatorial torus could be ejected, while the enhanced magnetic field could 
drive a collimated wind. The latter scenario could produce a configuration with an equatorial torus 
and collimated winds as observed in IRAS 17347$-$3139.
  
It has been estimated that the outflow driven by the dynamo action would be explosive and last
$\lesssim$ 100 years (Nordhaus \& Blackman 2006). This value is similar to the kinematical age of 
the ionized component of IRAS 17347$-$3139. However, this age represents only a lower limit for 
the age of the bipolar lobes, suggesting that the collimated wind has been present for more than 
100 years. Furthermore, given the low percentage of polarization derived from our OH 1612 MHz maser 
observations (see \S3.2), it is probable that the magnetic field is not very strong. This results 
suggest that the magnetic fields could not be playing a mayor role in the formation of the collimated 
structure of this PN. More measurements of the strenght and geometery of the magnetic field are required to 
further test this model.  

Another mechanism, proposed to explain the formation of collimated winds and the ejection of equatorial 
tori, also assumes that a binary companion spirals-in through the CSE of the AGB star, providing energy 
to detach the torus. When the spiral-in process of the secondary stops it could undergo Roche lobe 
overflow to form an accretion disk around the primary star. Alternatively, the companion could reach a region
 where it is shredded by gravitational tidal forces to form an accretion disk. Therefore, in a similar
 way to the star-forming regions, these disks may blow collimated winds that shape the PNs. These modeles
 predict the presence of an equatorial torus and a colimated wind. They also predict that the ejection of the 
equatorial torus precedes the formation of the jets. For the case of the PN IRAS 17347$-$3139, as 
mentioned above, we estimate that the age of the collimated wind must be greater than 100 years. 
In addition, from our high resolution image at 0.7 cm (Fig 8), we can estimate an inner radius 
for the torus, which is of the order of half the separation between the two intensity peaks 
($\sim 0\rlap{.}^{\prime\prime}06$); for a distance range of 0.8-6 kpc, and assuming a typical 
expansion velocity of 10 km~s$^{-1}$ (Huggins 2007), we find that the torus was completely ejected 
$\sim$25-185 years ago. To determine if the the torus was ejected previously than the collimated wind, 
and thus be able to test these models, we need a more accurate estimation of the distance to this 
source.

\section{Conclusions}

We have carried out sensitive high angular resolution VLA observations of the young 
planetary nebula IRAS 17374$-$3139. We present the first images of its ionized structure at 
cm wavelengths. The radio continuum images revealed the presence of a bright central structure, 
and an extended more tenuous bipolar component. 

A double Gaussian fit shows that the extended component is elongated in the same direction as the bipolar 
lobes observed in the near-infrared images, while the central structure shows an elongation in the 
perpendicular direction, parallel to the dark lane observed in the IR images. We interpret that the radio
continuum emission is arising in two extended ionized lobes, and in an equatorial ionized torus. The 
electron density at the base of the lobes is 2-6$\times$10$^{5}$~cm$^{-3}$, for a distance range from 6 
to 0.8 kpc. This relatively high density in the lobes supports the idea that this source is a very young
 PN. Given the subtle point-symmetric morphology of the extended component, we suggest the possible 
presence of a collimated ionized wind in this source. On the other hand, we derive a lower limit for 
the electron density for the equatorial torus of $n_{e} \geq 1$-$3\times10^6$~cm$^{-3}$, given the same 
distance range as above. A high resolution image at 0.7 cm reveals the presence of a double peak 
structure in the central component, supporting the interpretation of the equatorial torus. We compared 
the distribution of the water maser emission with our high resolution radio continuum images; the 
relative positions of the maser and the continuum emission suggest that the water masers arise from the
 outer parts of the ionized torus.

We detected OH maser emission at 1612~MHz toward IRAS~17347$-$3139. The spectrum shows only one weak 
feature at $V_{\rm LSR}=$~$-$70~km~s$^{-1}$ which coincides spatially with the continuum emission. 
We derived a 3-$\sigma$ upper limit of $<$ 35\% for the percentage of circularly polarized emission 
(m$_{c}$=V/I). We also report the detection of OH 1612 MHz maser emission coming from two other 
sources, J17380406-3138387 and OH 356.65-015, located within our primary beam.

\acknowledgments

DT and YG acknowledge support DGAPA-UNAM grant IN100407 and CONACyT grant 49947. GA, JFG 
and JMT acknowledge support from MEC (Spain) grant AYA2005-08523-C03 (co-funded with FEDER funds).
 LFM is supported by MEC grant AYA2005-01495 (co-funded with FEDER funds). GA, IdG, JFG, JMT and 
LFM are also supported by Consejer\'{\i}a de Innovaci\'on, Ciencia y Empresa of Junta de 
Andaluc\'{\i}a. We thank the anonymous referee for constructive comments on the manuscript.



\clearpage

\clearpage

\clearpage

\clearpage

\begin{figure*}
\begin{center}
\vspace{3cm}
\includegraphics[angle=-90,scale=0.7]{f4.eps}
 \caption{Contours: VLA image of IRAS 17347$-$3139 at wavelength 0.7 cm using a ROBUST weight parameter 
$= 0$. Contours are $-$2.5, 2.5, 3, 5, 10, 20, 30, 40, 50, 60, 70, 80, 90, 99 per
 cent of 7.35$\times10^{-2}$ Jy/beam, the peak value of the brightness.  The value of the first 
contours are -3 and 3 times the $rms$ noise of the image, 6.2 $\times 10^{-4}$ Jy/beam. The 
synthesized beam is shown in the bottom right corner, and its size is $0\rlap{.}^{\prime\prime}14
 \times 0\rlap{.}^{\prime\prime}10$  
(PA = $46^\circ$). Grey scale: HST IR image of IRAS 17347$-$3139 at 1.1 $\mu$m obtained with 
the NIC1 in the F110W filter. 
\label{fig_4}}
\end{center}
\end{figure*}

\clearpage

\begin{figure}
\begin{center}
\vspace{3cm}
\includegraphics[angle=0,scale=0.75]{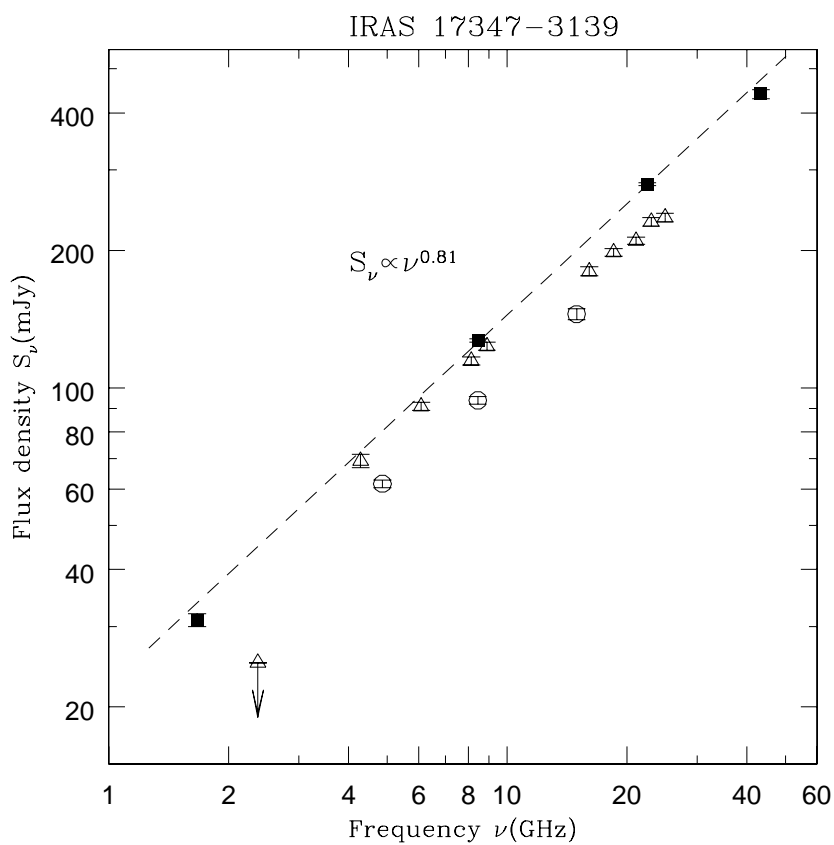}
\caption{Spectral energy distribution of IRAS~17347$-$3139 in the range $\sim$ 1-43 GHz. 
The filled squares represent the flux densities obtained from our observations [epoch 2005.1]. The 
open triangles come from the observations carried out by G05 [epoch 2004.2]. The open circles come 
from the observations carried out by dGM04 [epoch 2002.5]. Note the increasing trend of the flux 
density of the source with time. The dashed line is a linear fit to the data from our 
observations. 
\label{fig_5}}
 \end{center}
\end{figure}

\clearpage

\begin{figure}
\begin{center}
\includegraphics[angle=0,scale=0.45]{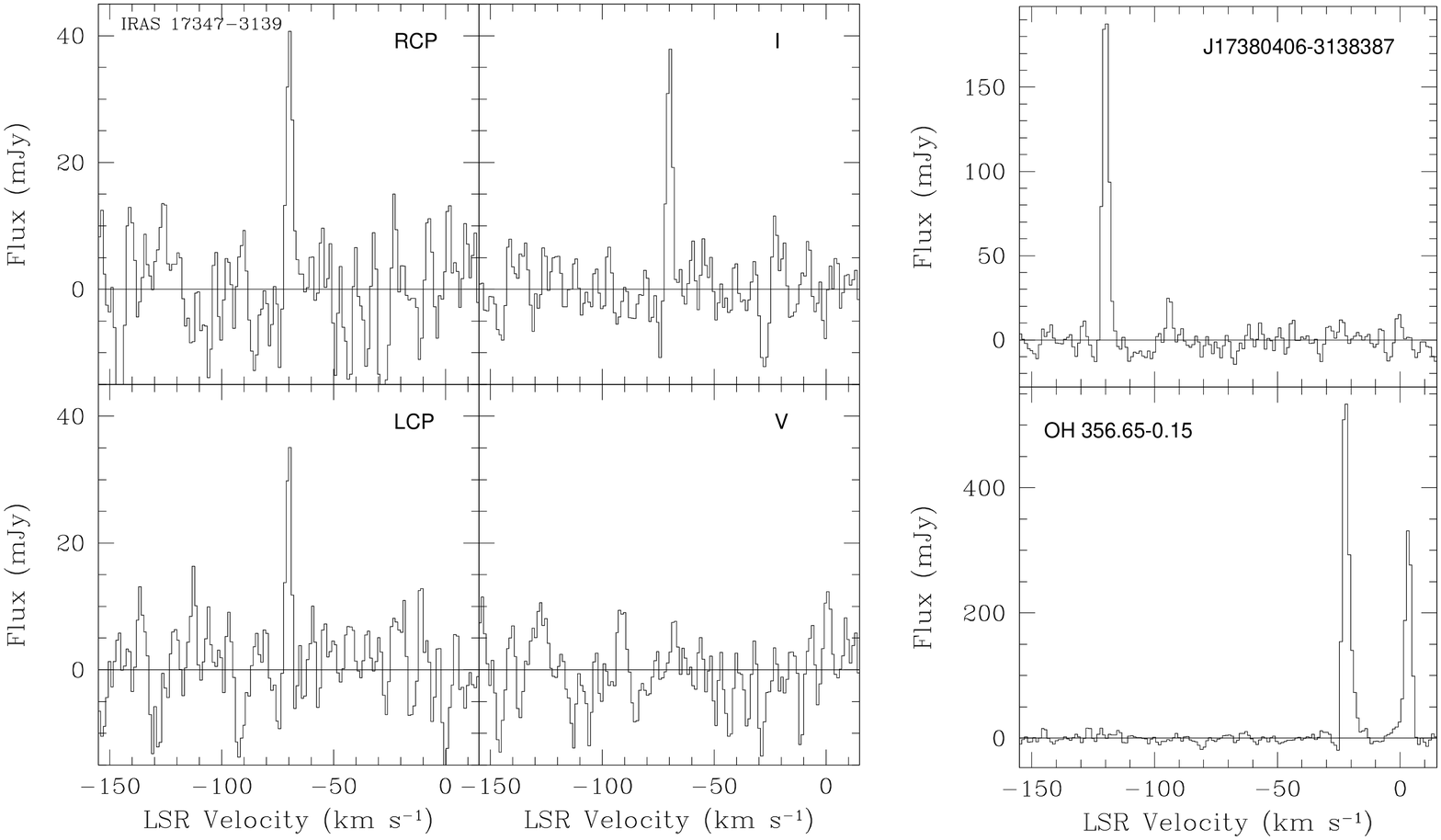}
\caption{Left panel: Spectra of the right and left circular polarizations, as well as 
those of the total flux density (I) and polarization (V) of the OH~1612~MHz maser emission 
toward IRAS~17347$-$3139. Right panel: OH~1612~MHz maser emission from two other sources in the field 
(see \S 3.2).
\label{fig_6}}
 \end{center}
\end{figure}

\begin{figure}
\begin{center}
\includegraphics[angle=0,scale=0.8]{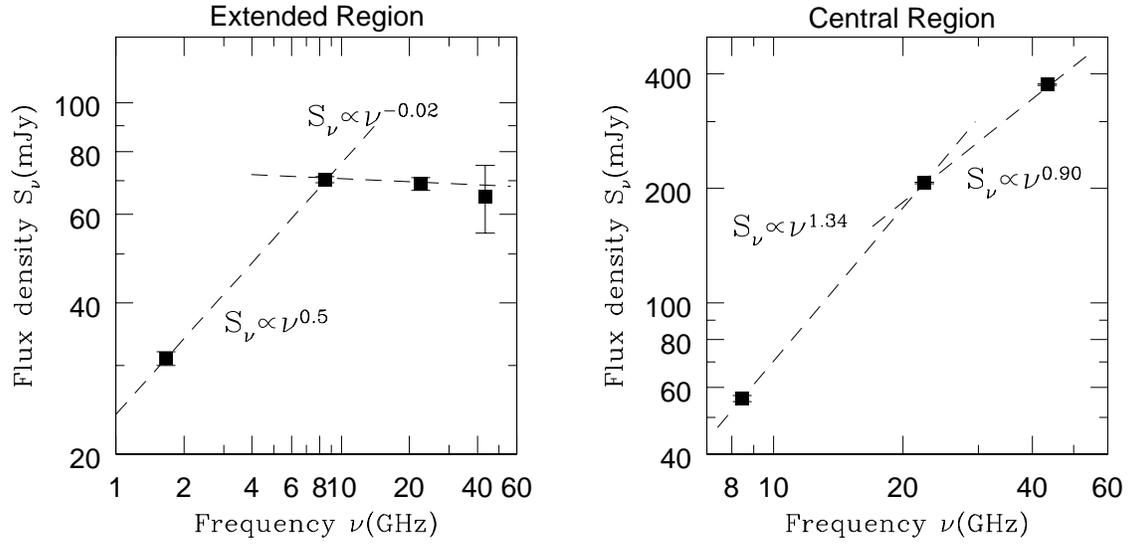}
\caption{Left: Spectral energy distribution of the extended emission in  IRAS~17347$-$3139.  
Right: Spectral energy distribution of the central component in IRAS~17347$-$3139.  
\label{fig_7}}
 \end{center}
\end{figure}

\clearpage

\begin{figure}
\begin{center}
\vspace{3cm}
\includegraphics[angle=-90,scale=0.7]{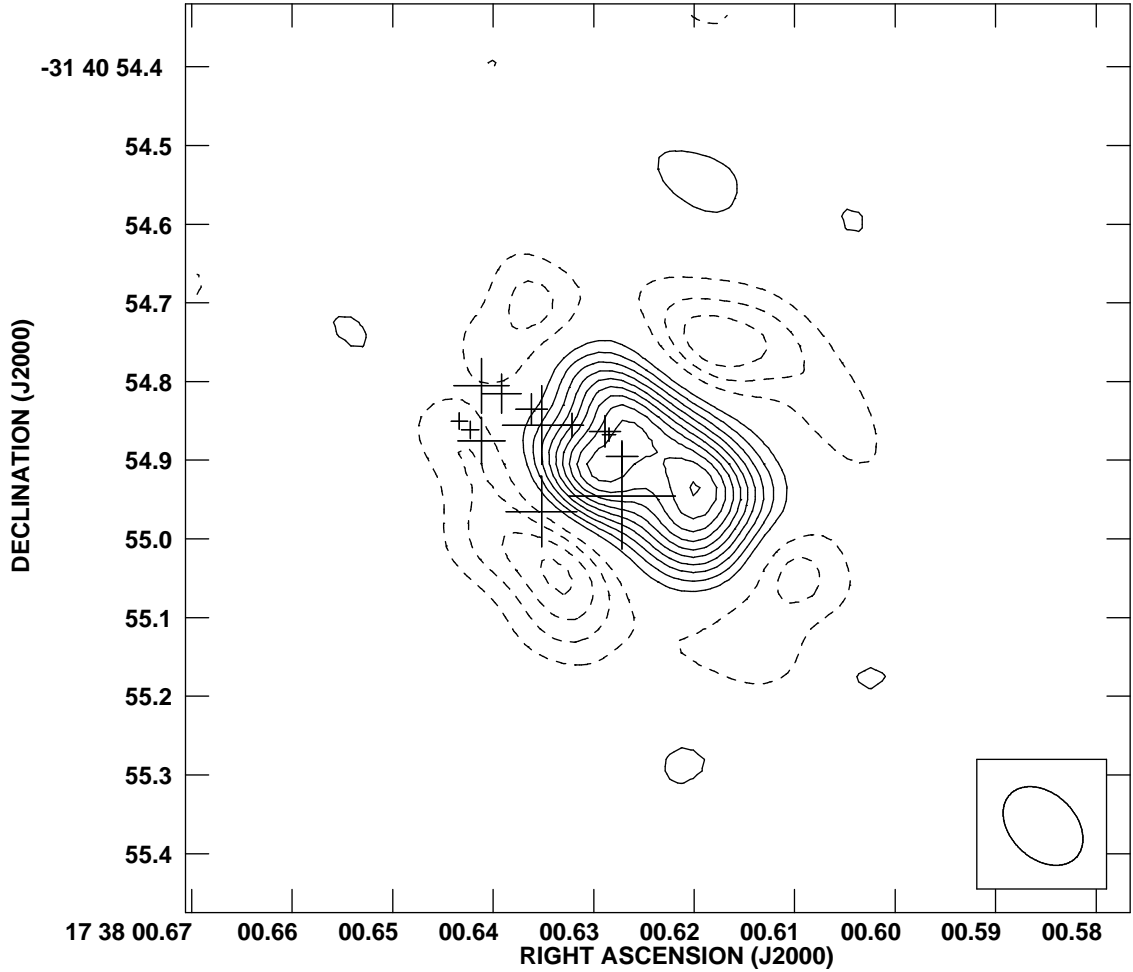}
\caption{VLA image of IRAS 17347$-$3139 at wavelength 0.7 cm using  a ROBUST weight parameter 
$= 0$, and restricting the visibilities to those with a $uv$ baseline longer than 500 k$\lambda$. 
Contours are $-$40, $-$30,$-$20,$-$10, 10, 20, 30, 40, 50, 60, 70, 80, 90, 99 per cent of
 2.6 $\times 10^{-2}$ Jy/beam, the peak value of the brightness. The value of the first contour is 
$-3$ and 3 times the $rms$ noise of the image, 7.4~$\times 10^{-4}$~Jy/beam.  The synthesized 
beam is shown in the bottom right corner, and its size is $0.12'' \times 0.08''$  (PA = $46^\circ$). 
Superimposed are shown the positions of the masers reported by dGM04.
\label{fig_8}}
 \end{center}
\end{figure}



\clearpage

%
%

\input{tab1.tex}
\input{tab2.tex}
\input{tab3.tex}
\input{tab4.tex}

\end{document}

%% file: tab1.tex
\begin{deluxetable}{rcrcr}
\tabletypesize{\tiny}
\tablecolumns{5} 
\tablecaption{Set up of the observations \label{table_1}}
\tablewidth{0pc}
\tablehead{
\colhead{Frequency} & \colhead{Phase Calibrator} & \colhead{Flux Density\tablenotemark{a}} &\colhead{Flux Calibrator}& \colhead{Flux Density\tablenotemark{b}}\\
 \colhead{(GHz)} & & \colhead{(Jy)} & & \colhead{(Jy)} }
\startdata
 1.612 & J1751-253 & 0.94 $\pm$ 0.01& 3C 286 & 13.85\phantom{000}\\
 1.665 & ''        & 1.04 $\pm$ 0.01& ''     & 13.63\phantom{000}\\
 1.667 & ''        & 1.05 $\pm$ 0.01& ''     & 13.62\phantom{000}\\
 1.720 & ''        & 1.03 $\pm$ 0.01& ''     & 13.41\phantom{000}\\
 8.460 & J1744-312 & 0.78 $\pm$ 0.01& ''     &  5.21\phantom{000}\\
22.460 & ''        & 0.91 $\pm$ 0.02& ''     &  2.52\phantom{000}\\
43.340 & ''        & 1.08 $\pm$ 0.06& ''     &  1.45\phantom{000}
\enddata

\tablenotetext{a}{Bootstrapped flux density for phase calibrator.}
\tablenotetext{b}{Assumed flux density for flux calibrator.}

\end{deluxetable}

%% file: tab2.tex
\begin{deluxetable}{rccccccc}
\tabletypesize{\tiny}
\tablecolumns{8} 
\tablecaption{Parameters of the gaussian fitting to the continuum emission of IRAS~17347-3139\tablenotemark{a}\label{table_2}}
\tablewidth{0pc}
\tablehead{
\multicolumn{1}{c}{}&
\multicolumn{3}{c}{Compact region}&
\multicolumn{3}{c}{Extended region}\\
\multicolumn{1}{c}{}&
\multicolumn{3}{c}{\hrulefill}&
\multicolumn{3}{c}{\hrulefill}\\
\colhead{Frequency} & \colhead{Flux density} & \colhead{Size\tablenotemark{b}} & \colhead{P.A.}& \colhead{Flux density} & \colhead{Size\tablenotemark{b}} & \colhead{P.A.}\\
\colhead{(GHz)} &  \colhead{(mJy)} & \colhead{(arcsec $\times$ arcsecs)} & \colhead{(degrees)}& \colhead{(mJy)} &\colhead{(arcsec $\times$ arcsecs)} & \colhead{(degrees)}
}
\startdata
 1.666\tablenotemark{c} &  $\ldots$ & $\ldots$ & $\ldots$ & 31$\pm$1& 2.20$\times$0.40&$-$29.8\\
 8.460 &  \phantom{0}56$\pm$1 & 0.33$\times$0.19&16.1&70$\pm$1&1.57$\times$0.40&$-$32.9\\
22.460 &  207$\pm$1& 0.25$\times$0.22&57.8&69$\pm$2 &1.42$\times$0.47&$-$32.7\\
43.340 &  375$\pm$2& 0.25$\times$0.19& 50.4& 65$\pm$10\tablenotemark{d}& $\ldots$&$\ldots$
\enddata
%
%
\tablenotetext{a}{Two gaussian components fitting, except for the data at 1.666 and 43.340 GHz.}
\tablenotetext{b}{Deconvolved size.}
\tablenotetext{c}{Average frequency of the four OH transitions (see \S 2).}
\tablenotetext{d}{Residual flux density from the single Gaussian fit of the compact region (see \S 3.1).}
\end{deluxetable}

%% file: tab3.tex
\begin{deluxetable}{rcccc}
\tabletypesize{\tiny}
\tablecolumns{5} 
\tablecaption{Continuum emission of IRAS~17347-3139\label{table_3}}
\tablewidth{0pc}
\tablehead{
\colhead{Frequency} & \colhead{Flux density\tablenotemark{a}}& \colhead{RA(J2000)\tablenotemark{b}} & \colhead{Dec(J2000)\tablenotemark{b}}& \colhead{Position Uncertainty\tablenotemark{c}}\\
\colhead{(GHz)} & \colhead{(mJy)}& \colhead{(h m s)} & \colhead{$\left( ^{\circ\;\;\prime\;\;\prime\prime}\right)$} & \colhead{(arcsec)}
}
\startdata
 1.666\tablenotemark{d} & \phantom{0}31 $\pm$  1 & 17 38 00.61 & $-$31 40 55.0 & 0.4\\
 8.460 &           127 $\pm$  1 & 17 38 00.63 & $-$31 40 54.9 & 0.1\\
22.460 &           280 $\pm$  2 & 17 38 00.624 & $-$31 40 54.90 & 0.05\\
43.340 &           440 $\pm$ 10 & 17 38 00.624 & $-$31 40 54.91 & 0.05
\enddata
%
%
\tablenotetext{a}{Total flux density of the emission. The uncertainties were obtained using 1 $\sigma$ the $rms$ noise of the image.}
\tablenotetext{b}{Position of the emission peak obtained from a Gaussian fitting.}
\tablenotetext{c}{Absolute position error.}
\tablenotetext{d}{Average frequency of the four OH transitions (see \S 2).}
\end{deluxetable}

%% file: tab4.tex
\begin{deluxetable}{cccccc}
\tabletypesize{\tiny}
\tablecolumns{6} 
\tablecaption{OH 1612 MHz maser emission\tablenotemark{a}\label{table_4}}
\tablewidth{0pc}
\tablehead{
\colhead{Source name} & \colhead{RA(J2000)\tablenotemark{b}} & \colhead{Dec(J2000)\tablenotemark{b}} & \colhead{LSR Velocity range} & \colhead{Flux density\tablenotemark{c}} & \colhead{rms noise}  \\
&\colhead{(h m s)} & \colhead{$\left( ^{\circ\;\;\prime\;\;\prime\prime}\right)$} &\colhead{(km s$^{-1}$)} & \colhead{(mJy)} & \colhead{(mJy/beam)\tablenotemark{d}} 
}
\startdata
IRAS 17347$-$3139 &17 38 00.57  & $-$31 40 54.9  & $-$$70 $ &  38 $\pm$ 7   & 4 \\
OH 356.65$-$0.15    & 17 38 00.66 & $-$31 51 54.4  &\phantom{0}3 &    335 $\pm$ 7  & 4  \\
                    & 17 38 00.65 & $-$31 51 54.4  &   $-$22     &    543 $\pm$ 7  & 4  \\
J17380406$-$3138387  & 17 38 04.10 & $-$31 38 38.3  &   $-$95     &     30 $\pm$ 7  & 4  \\
                    & 17 38 04.06 & $-$31 38 38.6  &   $-$123    &    200 $\pm$ 7  & 4 
\enddata
\tablenotetext{a}{No emission has been detected from the 1665, 1667 and 1720 MHz OH maser transitions. The $rms$ noise 
is 4 mJy/beam for all the transitions.}
\tablenotetext{b}{Position of the emission peak of the observed spectral feature obtained from a Gaussian fitting. The absolute position error is 0$\rlap{.}^{\prime\prime}$4}
\tablenotetext{c}{Peak flux density of the observed spectral feature (The integration region was 6$\rlap{.}^{\prime\prime}$5 $\times$ 5$\rlap{.}^{\prime\prime}$5 for IRAS 17347$-$3139, and 8$^{\prime\prime}$$\times$8$^{\prime\prime}$ for OH 356.65$-$0.15 and J17380406$-$3138387).  The uncertainties were obtained using 1 $\sigma$ the $rms$ noise of the image.} 
\tablenotetext{d}{1 $\sigma$ the $rms$ noise per channel. The size of the synthesized beam is 4$\rlap{.}^{\prime\prime}$4 $\times$ 2$\rlap{.}^{\prime\prime}$8 (Position angle = 65 degrees)}

\end{deluxetable}